\documentclass[a4paper,notitlepage,tightenlines,twocolumn,nofootinbib,superscriptaddress,preprintnumbers,showpacs]{revtex4-2}

\usepackage{times}
\usepackage{mathptmx}   
\DeclareMathAlphabet{\mathcal}{OMS}{cmsy}{m}{n} 

\usepackage{amssymb,amsbsy,amsmath,amsfonts}
\usepackage{mathrsfs}

\usepackage{graphicx}
\usepackage{graphics,psfrag}

\usepackage{setspace}

\usepackage{nicefrac}

\usepackage{esint}

\usepackage{natbib}

\usepackage{graphicx}
\usepackage{epsf,epsfig,float,latexsym,amsthm,fancyhdr,rotating}
\usepackage{graphics,psfrag,longtable}
\usepackage{slashed}
\usepackage[normalem]{ulem}
\usepackage[colorlinks,citecolor=blue,linktoc=all,linkcolor=cyan,urlcolor=blue]{hyperref}
\usepackage{upgreek}

\def\beq{\begin{equation}}
\def\eeq{\end{equation}}
\def\bea{\begin{eqnarray}}
\def\eea{\end{eqnarray}}
\def\beqa{\begin{equation}\begin{array}{l}}
\def\eeqa{\end{array}\end{equation}}
\def\eqlab#1{\label{eq:#1}}

\def\Eqref#1{Eq.~(\ref{eq:#1})}

\def\barr{\left(\begin{array}{c}}
\def\earr{\end{array}\right)}
\def\bmat{\left(\begin{array}{cc}}
\def\emat{\end{array}\right)}
\def\al{\alpha}

\def\ga{\gamma}

\def\dd{\mathrm{d}}

\def\3d{3-D}

\def\piEFT/{$\slashed{\pi}$EFT}

\usepackage{xcolor}
\usepackage{bm}
\usepackage{slashed}
\usepackage{graphicx}
\allowdisplaybreaks

\makeatletter
\g@addto@macro\bfseries{\boldmath}
\makeatother

\begin{document}
\preprint{MITP-22-050}
\preprint{PSI-PR-22-18}

\author{Vadim Lensky}
\affiliation{Institut f\"ur Kernphysik,
 Johannes Gutenberg-Universit\"at  Mainz,  D-55128 Mainz, Germany}

 \author{Franziska Hagelstein}
\affiliation{Institut f\"ur Kernphysik,
 Johannes Gutenberg-Universit\"at  Mainz,  D-55128 Mainz, Germany}
\affiliation{Paul Scherrer Institut, CH-5232 Villigen PSI, Switzerland}

\author{Vladimir Pascalutsa}
\affiliation{Institut f\"ur Kernphysik,
 Johannes Gutenberg-Universit\"at  Mainz,  D-55128 Mainz, Germany}

\title{A reassessment of nuclear effects in muonic deuterium using pionless effective field theory at N3LO}

\begin{abstract}
We provide a systematic assessment of the order-$\alpha^5$ nuclear contributions to the Lamb shift
 of muonic deuterium, including the accompanying radiative corrections due to vacuum polarization, up to next-to-next-to-next-to-leading order (N3LO) within the pionless effective field theory (\piEFT/). We also evaluate higher-order corrections due to the single-nucleon structure, which are expected to be the most important corrections beyond N3LO. We find a correlation between the deuteron charge and Friar radii, which can be useful to judge the quality of charge form factor parametrisations.
We refine the theoretical description of the $2\gamma$-exchange contribution, especially in the elastic contribution and the radiative corrections, ameliorating the original discrepancy between theory and experiment in the size of $2\gamma$-exchange effects. Based on the experimental Lamb shift of muonic deuterium, we obtain the deuteron charge radius, $r_d(\mu\text{D})=2.12763(13)_\text{exp}(77)_\text{theory}$~fm, which is consistent with (but less precise than) the value obtained by combining the H-D isotope shift with the muonic hydrogen Lamb shift. The theory uncertainty is evaluated using a Bayesian procedure and is dominated by the truncation of the \piEFT/ series.
\end{abstract}

\date{\today}

\maketitle


\section{Introduction}
Recent advances in the spectroscopy of light muonic atoms, by the CREMA Collaboration at the Paul Scherrer Institute, offer a new leap in precision for studies of nucleon and nuclear structure \cite{Antognini:2021icf}. 
Their determinations of the charge radii of the proton, $r_p$, and the deuteron, $r_d$, from the
Lamb shift of muonic hydrogen ($\mu$H) \cite{Pohl:2010zza,Antognini:1900ns} and muonic deuterium ($\mu$D) \cite{Pohl1:2016xoo}, have been particularly intriguing.
The former disagreed, quite spectacularly, with the previous determinations of $r_p$ using the conventional methods of hydrogen (H) spectroscopy and electron-proton ($ep$) scattering, \textit{viz.}, the ``proton-radius puzzle'' (see, e.g., \cite{Carlson:2015jba,Gao:2021sml}). The $\mu$D determination of  $r_d$ disagrees with CODATA-2014 \cite{Mohr:2015ccw}, but also
with a more recent determination from deuterium (D) \cite{Pohl:2016glp}. Initially, it even disagreed with $\mu$H, given the precise determination of the radius difference, $r_d^2- r_p^2$, from the H-D isotope shift. The latter discrepancy has recently been resolved on the theory side
by adding several missing contributions to the $\mu$D Lamb shift~\cite{Pachucki:2018yxe,Kalinowski:2018rmf}: most notably, electron vacuum polarization (eVP) corrections to the two-photon exchange ($2\ga$ exchange), as well as a  three-photon-exchange ($3\ga$-exchange) contribution.
These are certainly interesting improvements, motivating a more systematic account of nuclear effects
in $\mu$D and light muonic atoms in general. 

In this work, we give a reassessment of the finite-size and polarizability effects in $\mu$D
using the pionless effective field theory ($\slashed{\pi}$EFT) of nuclear forces \cite{Kaplan:1996xu,Kaplan:1998we,Kaplan:1998tg,Chen:1999bg,Chen:1999tn,Rupak:1999rk,Phillips:1999hh} at next-to-next-to-next-to-leading order (N3LO), including effects generated by the structure of individual nucleons.
The \piEFT/ framework gives a better control of theoretical uncertainties
in a well-defined perturbation theory, with the small parameter given by $P/m_\pi\sim\gamma/m_\pi
\simeq 1/3$, where the typical momentum scale $P$ in the deuteron is characterized by
the binding momentum $\gamma = \sqrt{M_N B}\simeq 45$ MeV, with $M_N$ the nucleon mass, $B$ the deuteron binding energy, and $m_\pi\simeq 139$~MeV the pion mass.
To quantify the uncertainty due to omitted higher-order terms in the \piEFT/ expansion, we use the methods of Bayesian statistics  along the lines of Refs.~\cite{Furnstahl:2015rha,Perez:2015ufa}.
The momentum scale probed in $\mu$D is of the order of $ \al\,m\lesssim 1$~MeV (with $\al$ the fine-structure constant and $m$ the muon mass), well below the limiting scale of \piEFT/, set by $m_\pi$. The expansion in the atomic momentum appears naturally as part of the usual expansion in $\alpha$.

Our present work is based on the recent N3LO evaluation of the deuteron charge form factor (FF) and the forward longitudinal amplitude of doubly-virtual Compton scattering (VVCS) \cite{Lensky:2021VVCS}, from which we 
derive all of the nuclear effects at $O(\alpha^5)$, as well as their interference with 
the eVP.  At N3LO, all of the nuclear contributions are governed by only one free parameter --- the low-energy constant $l_1^{C0_S}$, which describes the coupling of the two-nucleon system to a Coulomb photon, and will be determined here
 very precisely from the isotope shift. The nuclear corrections to the $\mu$D Lamb shift are then
a free-parameter-free prediction of N3LO $\slashed{\pi}$EFT.

\begin{figure}[tbh]
\includegraphics[width=\columnwidth]{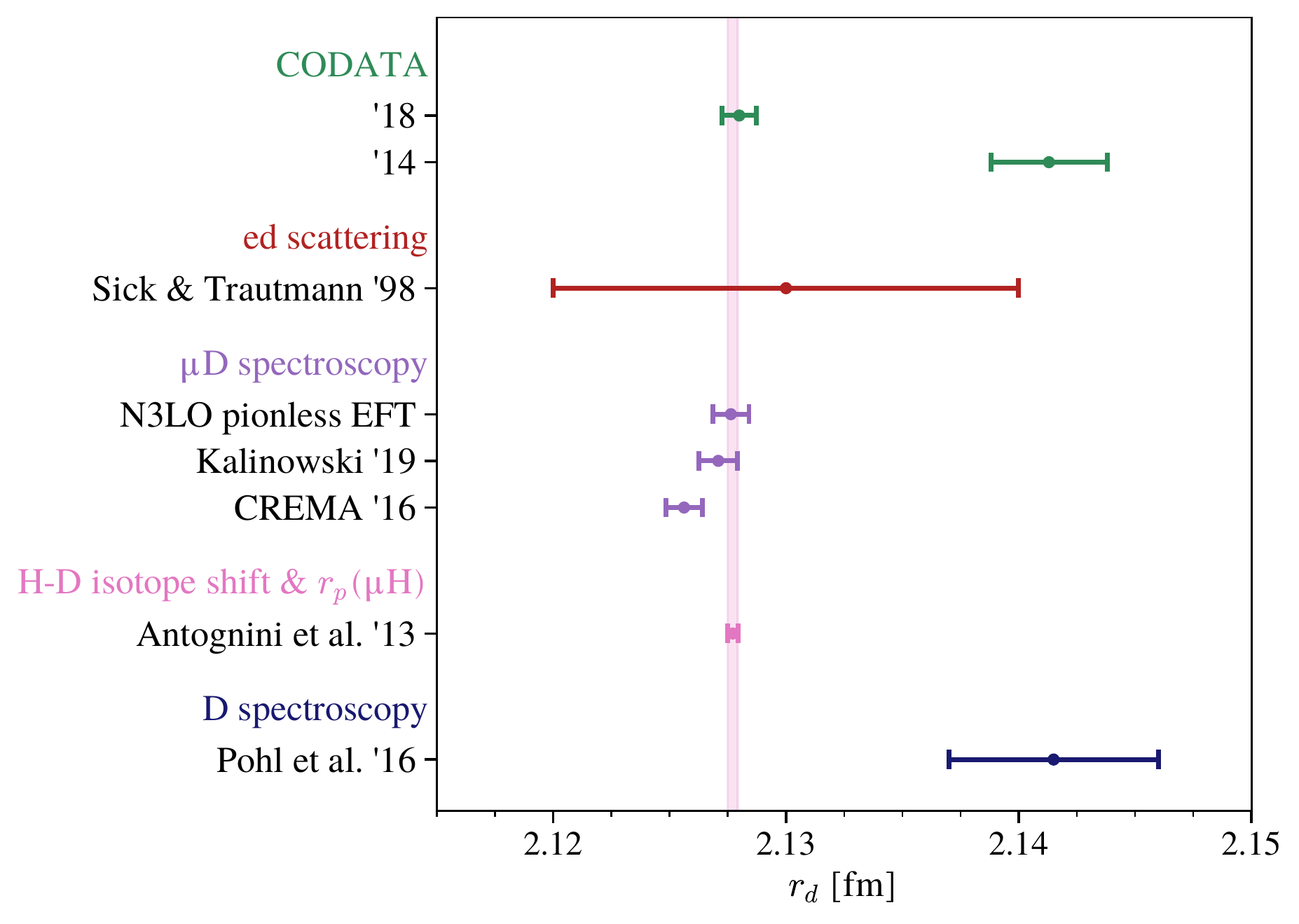}
\caption{Comparison of deuteron charge radius determinations from fits to electron-deuteron ($ed$) scattering data, the Lamb shift of muonic deuterium,  the $1S-2S$ hydrogen-deuterium isotope shift combined with the proton radius from muonic hydrogen, and deuterium spectroscopy. \label{fig:RdSummary}}
\end{figure}

Already at $O(\alpha^5)$, we find an appreciable difference with previous evaluations of the ``elastic'' $2\gamma$ contribution to the $\mu$D Lamb shift.
We then combine our results with the known 
$O(\alpha^6 \log \al ) $ and $O(\al^6)$ corrections, coming from the Coulomb distortion \cite{Krauth:2015nja} and
the $3\ga$-exchange \cite{Pachucki:2018yxe}, to obtain the full nuclear-structure contribution to the $\mu$D Lamb shift. It results in 
a slightly different extraction of the deuteron radius from the experimental $\mu$D Lamb shift, see Fig.~\ref{fig:RdSummary}. 
This extraction is now indeed consistent with the extraction from the isotope shift, but the latter is, of course, more precise.
In what follows, we give more details of this analysis, with emphasis on novel aspects 
of the $2\ga$ contribution
in $\slashed{\pi}$EFT. The ``elastic'' and ``inelastic'' $2\gamma$ contributions are evaluated in Sections \ref{SecII} and \ref{SecIII}, respectively. Hadronic contributions beyond N3LO are discussed in Section \ref{SecIV}. The total results are compiled in Section \ref{SecV}. A summary is given in Section \ref{SecVI}.

\section{Charge radius and elastic $\mathbf{2\gamma}$ exchange}\label{SecII}

We start with the finite-size contributions to the $\mu$D Lamb shift at N3LO in $\slashed{\pi}$EFT. Note that at this order $\slashed{\pi}$EFT gives very simple results, many of which can be computed analytically.

The main nuclear-structure effect in the Lamb shift is the finite-size 
 contribution of $O(\al^4)$:
\beq 
E_{nS}^{\mathrm{f.s.} } =  \frac{2 \al^4 m_r^3}{3 n^3}  r_d^2, 
\eeq 
with $n$ the principal quantum number and $m_r=m M_d/(m + M_d)$ the atomic reduced mass. The squared charge radius, in the case of the deuteron, is given by the
slope of the charge FF: $r_d^2 = -6\, d G_C(Q^2)/dQ^2|_{Q^2\to 0} \equiv -6 G^\prime_C(0)$. The N3LO \piEFT/ result for $r_d^2$ reads, order-by-order:
\beq
  r_d^2  =
    \frac{1}{8 \gamma ^2}
    +\frac{Z-1}{8 \gamma ^2}
    +2r_0^2
    +\frac{3(Z-1)^3}{\gamma ^2}\,l_1^{C0_S}\,, \label{eq:rdl1}
\eeq
where
$Z=1.67893(30)$ is the residue of the $NN$ scattering amplitude at the deuteron pole~\cite{Phillips:1999hh}, and
$r_0^2 = \nicefrac{1}{2}\left[r_p^2 + \nicefrac{3}{4}\,M_p^{-2} + r_n^2 \right]$ is the isoscalar nucleon charge radius, with the proton Darwin-Foldy term $\nicefrac{3}{8}\,M_p^{-2}$ added to it.
 In Ref.~\cite{Lensky:2021VVCS}, $l_1^{C0_S}$ was chosen to reproduce the deuteron charge radius from $\mu$D spectroscopy~\cite{Pohl1:2016xoo}, $r_d=2.12562(78)\ \text{fm}$, resulting in $l_1^{C0_S}=-2.32(41)\times 10^{-3}\,,$
where the uncertainty in the brackets stems from the error of the deuteron radius and the uncertainty of $Z$. However, the extraction of $r_d^2$ from $\mu$D spectroscopy depends on the theory result for the $2\ga$-exchange correction (even though the contribution of $l_1^{C0_S}$ to the $2\ga$-exchange correction is small). To avoid this interdependence, it is reasonable to use the deuteron charge radius from the H-D $1S-2S$ isotope shift instead. The isotope shift also has a $2\ga$-exchange contribution, but it has a much smaller correlation with $r_d^2$, and the contribution of $l_1^{C0_S}$ to the isotope shift can safely be neglected at the current level of precision.
The value of $r_d$ obtained from the H-D isotope shift, using the $\mu$H value of $r_p$, is~\cite{Antognini:2012ofa}:
\begin{equation}
r_d(\mu\text{H, iso}) = 2.12771(22)~\text{fm},
\label{eq:rdiso}
\end{equation}
and the corresponding result for the electric contact term coupling is:
\begin{align}
    l_1^{C0_S}=-1.8(4)\times 10^{-3}\,.
\label{eq:contact_term_value2}
\end{align}
This value is used in the calculation of the $2\gamma$-exchange correction presented below.

\begin{figure}[ht]
\includegraphics[width=0.25\textwidth]{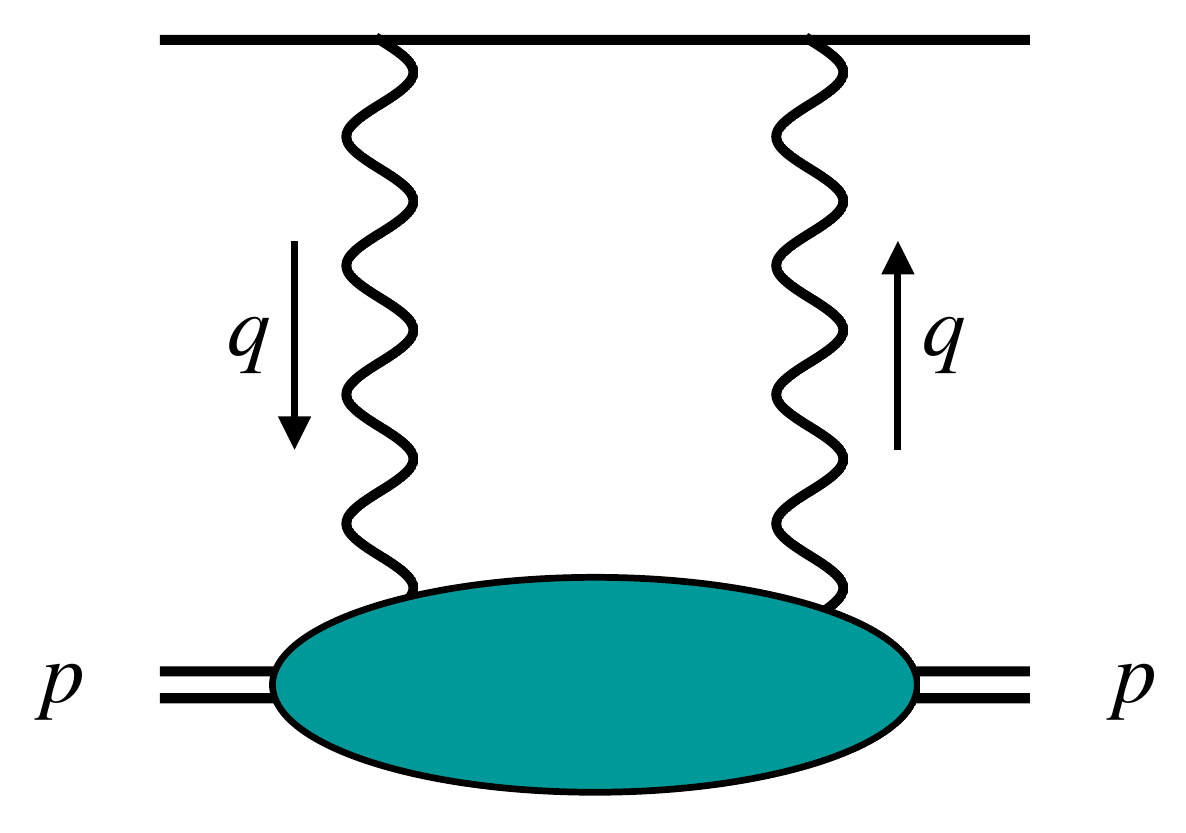}
\caption{The leading order in $\alpha$ two-photon-exchange potential. \label{fig:TPE}}
\end{figure}

\begin{figure}[htb]
\includegraphics[width=0.17\textwidth]{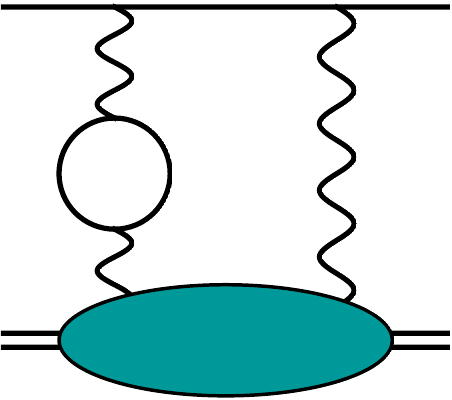}
\caption{Elastic and inelastic two-photon exchange with electronic vacuum polarization insertion at $O(\al^6)$. \label{fig:TPEwithVP}}
\end{figure}

The nuclear effects of $O(\al^5)$ can be described by the forward $2\ga$-exchange, see Fig.~\ref{fig:TPE}. 
There are also contributions involving eVP, shown in Fig.~\ref{fig:TPEwithVP}, which nominally are $O(\alpha^6)$, but, in the case of muonic atoms, are enhanced by factors of the muon-to-electron mass ratio $m/m_e$.

The forward $O(\alpha^5)$ $2\gamma$-exchange correction to the energy of a $nS$ state in $\mu$D is expressed through the forward VVCS off an unpolarised deuteron, which is parametrised by two scalar amplitudes $f_{L}(\nu,Q^2)$ and $f_T(\nu,Q^2)$ --- the longitudinal and transverse amplitudes, functions of the photon virtuality $Q^2$ and the lab-frame energy $\nu$~\cite{Drechsel:2002ar}. This $2\gamma$-exchange correction reads~\cite{Carlson:2013xea}:
\bea
   E_{nS}^\mathrm{fwd}&=&  8i\pi \al m \,\left[\phi_{n}(0)\right]^2\,
\int \!\!\frac{\dd^4 q}{(2\pi)^4}  \times\nonumber\\
& &\hspace{1.cm} \frac{f_L(\nu,Q^2)+2(\nu^2/Q^2)f_T(\nu,Q^2)}{Q^2(Q^4-4m^2\nu^2)},\label{eq:TPE_LT}
\eea
where $[\phi_{n}(0)]^2=1/(\pi n^3 a^3)$ is the (Coulomb) wave function of the $nS$ atomic state at the origin, $a=1/(\mathcal{Z} \al m_r)$ is the Bohr radius, and $\mathcal{Z} $ is the nuclear charge (equal to $1$ for the deuteron). The \piEFT/ counting assigns $Q=O(P)$, $\nu=O(P^2)$; the leading terms of $f_L$ and $f_T$ are, respectively, $O(P^{-2})$ and $O(P^0)$~\cite{Lensky:2021VVCS}. The factor $\nu^2/Q^2=O(P^2)$ in Eq.~\eqref{eq:TPE_LT} further suppresses the transverse contribution, which is N4LO relative to the leading longitudinal one. The N3LO result for $f_L(\nu,Q^2)$ obtained in Ref.~\cite{Lensky:2021VVCS} thus allows us to calculate the forward $2\gamma$-exchange correction up to N3LO in the \piEFT/ counting. 

The elastic contribution to the $2\ga$ exchange is given in terms of the elastic electromagnetic FFs~\cite{Carlson:2013xea}, and, neglecting the magnetic and quadrupole contributions (they both arise at higher orders in the \piEFT/ counting, and their smallness, $<1\,\upmu$eV, is confirmed using empirical deuteron FF parametrisations~\cite{Abbott:2000ak,Sick:1998cvq}), reads:
\begin{align}
E_{nS}^\mathrm{elastic}  = & -\frac{32m M_d \alpha^2}{M_d^2-m^2}[\phi_{n}(0)]^2\times\nonumber\\ 
 &\int\limits_0^\infty\frac{\mathrm{d}Q}{Q} 
\left[
\frac{G_C^2(Q^2)-1}{4Q^2}
\hat{\gamma}_2(\tau_d,\tau_l)
 +\frac{M_d-m}{Q}G_C'(0)
\right],\label{eq:contrib_elastic}
\end{align}
with the auxiliary functions  
$\hat{\gamma}_{2}(x,y) = \gamma_{2}(x)/\sqrt{x}-\gamma_{2}(y)/\sqrt{y}$,
and $\gamma_2(x)  = (1+x)^{3/2}-x^{3/2} -\nicefrac{3}{2}\sqrt{x}$, and the dimensionless quantities $\tau_d=Q^2/(4M_d^2)$ and $\tau_l=Q^2/(4m^2)$.
The N3LO \piEFT/ result for $E_{2S}^\mathrm{elastic}$ yields, order-by-order:
\begin{eqnarray}
E_{2S}^\mathrm{elastic} &=&[-0.2043-0.1582-0.0626-0.0213]~\mathrm{meV}\nonumber\\
                        &=& -0.4463(77) \text{ meV},
\label{eq:elastic_piEFT}
\end{eqnarray}
with the uncertainty dominated by the higher-order terms in the \piEFT/ expansion, and estimated following the Bayesian approach in Refs.~\cite{Furnstahl:2015rha,Perez:2015ufa}. This 
significantly deviates from the result obtained in~Ref.~\cite{Carlson:2013xea}, $E_{2S}^\mathrm{elastic}=-0.417(2)$~meV, which used the $t_{20}$ deuteron FF parametrisation of Abbott et al.~\cite{Abbott:2000ak}, also adopted in the recent Ref.~\cite{Acharya:2020bxf}. At the same time, the \piEFT/ result is confirmed using the recent chiral effective theory ($\chi$ET) fit for the deuteron charge FF~\cite{Filin:2019eoe,Filin:2020tcs}, with which we obtained $E_{2S}^\mathrm{elastic}=-0.4456(18)\text{ meV}$, with the uncertainty corresponding to the $\chi$ET uncertainty of the charge FF. These results, together with the results for the inelastic contribution, are summarised in Table~\ref{tab:results_comparison_parts}.

The fact that the two low-energy effective theories give coinciding results vindicates their choice as tools to investigate the low-momenta properties of the deuteron. It also appears that the discrepancy in $E_{2S}^\mathrm{elastic}$ between the EFTs and the Abbott parametrisation~\cite{Abbott:2000ak} is due to the poor low-$Q$ properties of the latter. To elucidate this issue, we consider the Friar radius $r_{\mathrm{F}d}$, given in terms of the charge FF as:
\begin{align}
   r_{\mathrm{F}d}^3  & = \frac{48}{\pi}\int\limits_0^\infty \frac{\dd Q}{Q^4}\left[G_C^2(Q^2)-1-2G'_C(0)\,Q^2\right].
\end{align}
This integral coincides, up to a constant factor, with the leading term in the expansion of $E_{2S}^\mathrm{elastic}$ in powers of $m$~\cite{Hagelstein:2015egb}.
Up to N3LO in \piEFT/, it can be calculated analytically, the N3LO result being:
\begin{eqnarray}
    r_{\mathrm{F}d}^3 & = & \frac{3}{80\gamma^3}\bigg\{
    Z \left[5 - 2 Z (1-2\ln 2)\right]\nonumber\\
    & &-\frac{320}{9} r_0^2\gamma ^2 \left[Z(1-4\ln 2)-2+2\ln 2\right]\nonumber\\
    & & +
    80 (Z-1)^3\, l_1^{C0_S}\bigg\}.
    \label{eq:Friar}
\end{eqnarray}
\begin{figure}[htb]
    \centering
    \includegraphics[width=\columnwidth]{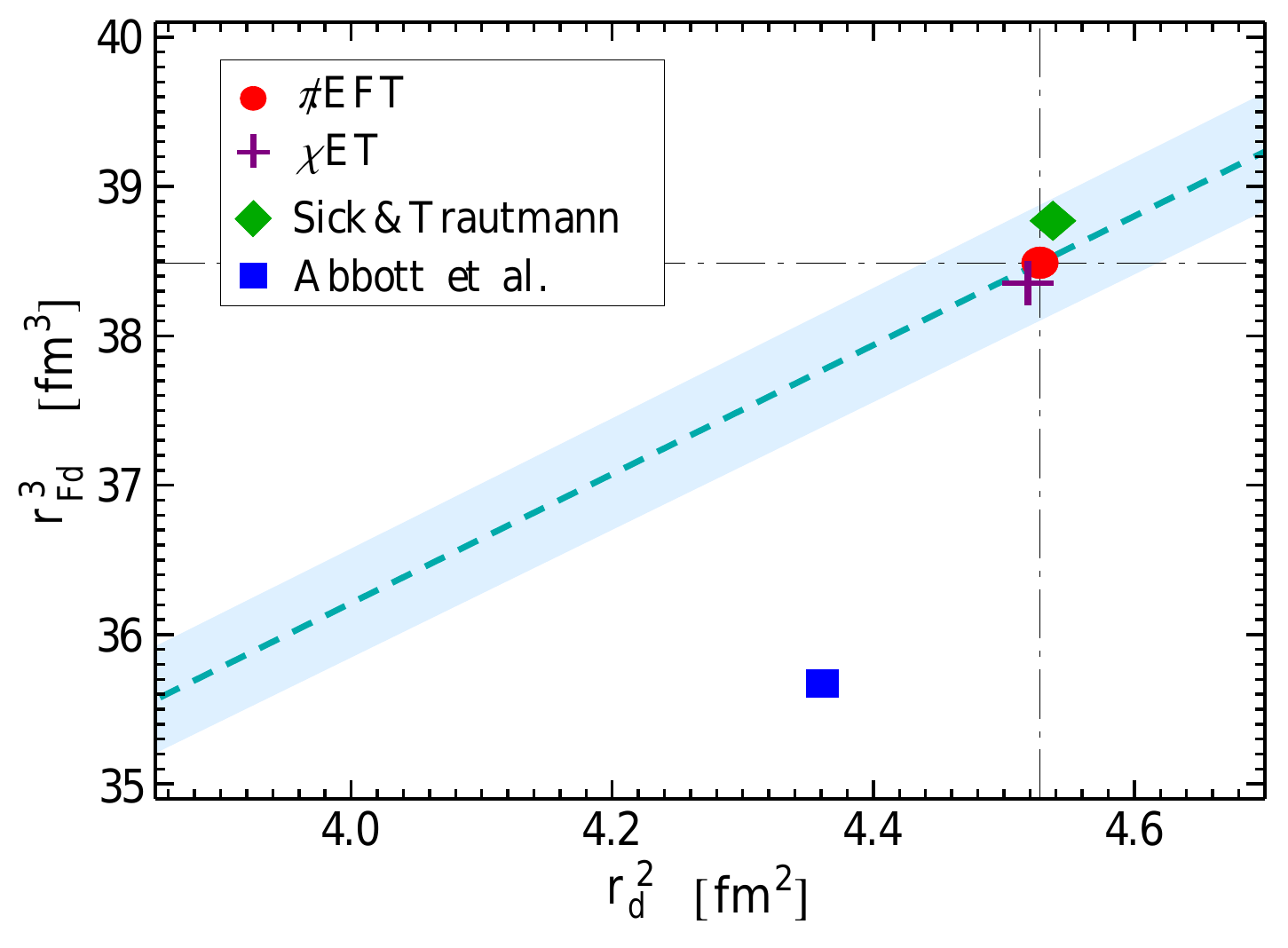}
    \caption{Correlation of $r_{\mathrm{F}d}^3$ and $r_d^2$. The dashed line shows the correlation obtained from the \piEFT/ results, with the band showing the estimated $1\%$ N3LO uncertainty. The red disc, purple cross, green diamond, and blue square show the values obtained, respectively, from \piEFT/, the $\chi$ET form factor~\cite{Filin:2020tcs}, the parametrisation of Ref.~\cite{Sick:1998cvq}, and the parametrisation of Ref.~\cite{Abbott:2000ak}. The dash-dotted lines indicate the isotope shift value of $r_d^2$ and the corresponding value of $r_{\mathrm{F}d}^3$ obtained at N3LO in \piEFT/.}
    \label{fig:RF_correlation}
\end{figure}

The dependence of both $r_d^2$, Eq.~\eqref{eq:rdl1}, and $r_{\mathrm{F}d}^3$, Eq.~\eqref{eq:Friar}, on $l_1^{C0_S}$ can be represented as a correlation line, which is shown in Fig.~\ref{fig:RF_correlation}, with the band corresponding to the estimated N3LO relative uncertainty of $1\%\simeq (\gamma/m_\pi)^4$. One can see that the point corresponding to the parametrisation of Abbott et al.~\cite{Abbott:2000ak} deviates from the correlation line, while the values calculated with the charge FF resulting from \piEFT/, the $\chi$ET fit of~\cite{Filin:2019eoe,Filin:2020tcs}, and the empirical parametrisation of~\cite{Sick:1998cvq} cluster very closely together. Note that it is not only the value of the deuteron charge radius in the parametrisation of~\cite{Abbott:2000ak}, $r_d=2.094(9)$, that tends to be lower than the other results, it is also the higher derivatives of the charge FF being smaller, resulting in a smaller $r_{\mathrm{F}d}^3$. The deuteron Friar radius as an integral low-$Q$ feature, and, in particular, its correlation with the deuteron charge radius, can be an important criterion for the quality of a FF parametrisation.

\section{Inelastic $\mathbf{2\gamma}$ Exchange} \label{SecIII}
The inelastic part of the $2\ga$-exchange contribution is given in terms of the non-pole parts of the VVCS amplitudes as:
\allowdisplaybreaks[0]
\begin{align*}
E_{nS}^\mathrm{inel}
=&-\frac{\alpha}{2\pi^2 m}\left[\phi_{n}(0)\right]^2\int\limits_0^\infty\frac{\mathrm{d}Q}{Q}\int\limits_{-1}^{1} \mathrm{d}x\,\sqrt{1-x^2}\,\times\nonumber\\
&\qquad\qquad\frac{f_L(-iQx,Q^2)-2x^2 f_T(-iQx,Q^2)}{
\tau_l+x^2},
\label{eq:LS_from_LT}
\end{align*}
which arises from Eq.~\eqref{eq:TPE_LT} after a Wick rotation and integration in hyperspherical coordinates. The respective pole parts, as well as the Thomson term entering $f_T$, are assumed to be subtracted from the VVCS amplitudes. The evaluation with the \piEFT/ VVCS amplitudes is straightforward, resulting in, order-by-order:
\begin{eqnarray}
E_{2S}^{\mathrm{inel},L}  &=& [-0.943 - 0.635 + 0.049 + 0.025]~\mathrm{meV}\nonumber\\
&=& - 1.504(16) \ \text{meV}
\label{eq:TPE_inelastic_orders}
\end{eqnarray}
for the longitudinal contribution, and
\begin{align}
E_{2S}^{\mathrm{inel},T}=-0.005\text{ meV}
\end{align}
for the LO$+$NLO transverse contribution. We include the latter in the final result despite its smallness, both in the \piEFT/ counting and numeric, for the sake of comparing with other calculations that typically include it. It is also in a very good agreement with the recent calculations of~Refs.~\cite{Acharya:2020bxf,Hernandez:2019zcm}. We therefore get:
\begin{equation}
    E_{2S}^\mathrm{inel}=E_{2S}^{\mathrm{inel},L}+E_{2S}^{\mathrm{inel},T}=-1.509(16)~\text{meV}.
\label{eq:E_inel}
\end{equation}
The uncertainty here is calculated in the same Bayesian approach as for $E_{2S}^\mathrm{elastic}$, with the uncertainty of the transverse part neglected.
The comparison of the results for $E_{2S}^\mathrm{inel}$ is shown in Table~\ref{tab:results_comparison_parts}. Our result agrees both with the recent covariant dispersive calculation~\cite{Acharya:2020bxf}, as well as with the calculation of Ref.~\cite{Hernandez:2019zcm}, within the uncertainties. The latter has a slightly larger in magnitude central value. The value obtained by us is appreciably smaller than the N2LO \piEFT/ result of Emmons et al.~\cite{Emmons:2020aov}, which is not unexpected, given that the nucleon-size effects, included in our calculation but not in Ref.~\cite{Emmons:2020aov}, suppress the magnitude of the $2\gamma$-exchange correction. The difference, however, is accommodated by the uncertainty of the N2LO calculation.
The data-driven evaluation of Carlson et al.~\cite{Carlson:2013xea} obtains an even larger $E_{2S}^\mathrm{inel}$, albeit with a large uncertainty making it compatible with any of the other results. 

{\renewcommand{\arraystretch}{1.5}
\begin{table}[t]
    \centering
    \begin{tabular}{l||c|c}
    Calculation & $E^\mathrm{elastic}_{2S}$ & $E^\mathrm{inel}_{2S}$ \\
    \hline\hline
    \piEFT/ N3LO & $-0.4463(77)$  & $-1.509(16)$ \\
      \piEFT/ p.N.\ N2LO \cite{Emmons:2020aov}~\footnote{Note that Ref.~\cite{Emmons:2020aov} uses a different variant of \piEFT/ and treats nucleons as point-like (p.N.). The value quoted is their ``$\mathcal{Z}_d$-improved'' result.}&& $-1.574(80)$\\
    $\chi$ET~\footnote{Evaluated using the form factors of Ref.~\cite{Filin:2020tcs}.} & $-0.4456(18)$  & \\
    \hline
    Carlson et al.~\cite{Carlson:2013xea} & $-0.417(2)$  & $-1.566(740)$ \\
    Acharya et al.~\cite{Acharya:2020bxf} &  & $-1.511(12)$ \\
    Hernandez et al.~\cite{Hernandez:2019zcm} & & $-1.531(12)$
    \end{tabular}
    \caption{Comparison of our results with other recent calculations for the elastic contribution $E^\mathrm{elastic}_{2S}$ and the inelastic contribution $E_{2S}^\mathrm{inel}$. Values are in meV. To compare with Ref.~\cite{Hernandez:2019zcm}, we subtract the subleading $O(\alpha^6\log\alpha)$ Coulomb correction from their ``$\eta$-less'' result. The uncertainty given here for their prediction is obtained based on the relative uncertainties of individual error sources from Ref.~\cite[Table 8]{Ji:2018ozm} (nuclear model, isospin symmetry breaking,
relativistic, higher $\mathcal{Z}\al$) summed in quadrature. }
    \label{tab:results_comparison_parts}
\end{table}}

\section{Hadronic contributions beyond N3LO} \label{SecIV}
In addition to the N3LO \piEFT/ result, we evaluate the corrections that arise due to the nucleon structure (such as the higher-order terms in the expansion of the nucleon FFs and the nucleon polarisabilities) at orders beyond N3LO. Those corrections are problematic in the strict \piEFT/ expansion, since they are leading to divergent contributions to $E_{nS}^\mathrm{fwd}$ and, as such, demand the inclusion of a two-nucleon-two-lepton contact term to regularise this divergence. We circumvent this problem, plugging in the full nucleon FFs in the nucleon-photon vertices at LO, similarly to what is done in $\chi$ET~\cite{Acharya:2020bxf,Filin:2020tcs}. This allows us to calculate the correction to the inelastic part of the $2\gamma$-exchange contribution due to higher-order terms in the nucleon FFs: 
\begin{align}
E _{2S}^\mathrm{hadr,\ FF} = -0.013(1)~\mathrm{meV}. \label{eq:HadrFF}
\end{align}
The respective effect on the elastic part of the $2\gamma$-exchange contribution is very small, and we neglect it.

The (inelastic and subtraction) contributions of the nucleon polarisabilities are calculated in a similar fashion. Namely, the inelastic contribution can be inferred from a dispersive integral with input from empirical deuteron structure functions at energies starting from the pion production threshold ~\cite{Carlson:2013xea}; on the other hand, both this term and the nucleon subtraction term can be calculated by re-scaling the sum of the respective proton and neutron contributions with the appropriate wave function factor~\cite{Krauth:2015nja}. For the inelastic contribution, we adopt the dispersive result of~\cite{Carlson:2013xea},
\begin{align}
E _{2S}^\mathrm{hadr,\ inel}=-0.028(2)~\mathrm{meV}.\label{eq:HadrInel}
\end{align}
Note that this result coincides with that of the re-scaling, using the single-nucleon values from~\cite{Tomalak:2018uhr}, which gives $-0.030(2)$~meV. For the subtraction contribution, we adopt the covariant chiral perturbation theory ($\chi$PT) result for the proton subtraction contribution~\cite{Lensky:2017bwi} and its neutron counterpart, and use the re-scaling procedure, getting
\begin{align}
E _{2S}^\mathrm{hadr,\ subt}=0.009(6)~\mathrm{meV},\label{eq:HadrSubt}
\end{align}
which agrees well with the value adopted in Ref.~\cite{Krauth:2015nja}: $0.0098(98)$~meV.

Adding these hadronic corrections together, we get
\begin{equation}
    E _{2S}^\mathrm{hadr} = E _{2S}^\mathrm{hadr,\ FF} + E _{2S}^\mathrm{hadr,\ subt} + E _{2S}^\mathrm{hadr,\ inel} =- 0.032(6)~\mathrm{meV}.
    \label{eq:higher_order_hadron_correction}
\end{equation}
Since the hadronic contribution is larger than our N3LO uncertainty estimate for $E_{2S}^\mathrm{inel}$, Eq.~\eqref{eq:E_inel}, we add it to the total. We expect these hadronic corrections to be the only sizeable contributions beyond N3LO, whereas purely nuclear effects at N4LO or higher in the \piEFT/ expansion should be fully contained by the given uncertainty estimate.

\section{Total $\mathbf{2\gamma}$ effect and the deuteron radius determination}\label{SecV}
To arrive at our final number for the nuclear-structure correction, we add two further contributions. The first of them is the eVP correction to the forward $2\gamma$ exchange, also calculated by us using the \piEFT/ VVCS amplitudes, 
\beq
    E _{2S}^\mathrm{eVP} =-0.027\text{ meV},\eqlab{eVP}
\eeq
with a negligibly small uncertainty. This result agrees with the first evaluation of this contribution by Kalinowski~\cite{Kalinowski:2018rmf}. The second contribution is the off-forward $2\gamma$ exchange (or Coulomb distortion). Despite being a subleading effect in the QED expansion, [$O(\alpha^6\log\alpha)$], it is needed for a meaningful extraction of $r_d$ from the $\mu$D Lamb shift and for a comparison with empirical results for the nuclear-structure effect. We adopt the value from the theory compilation~\cite{Krauth:2015nja},
\beq
E _{2S}^\mathrm{Coulomb}=0.2625(15)\,\text{meV}.\eqlab{Coulombdist}
\eeq
It is derived using modern nucleon-nucleon potentials; since the deuteron electric dipole polarizability obtained in \piEFT/ \cite{Lensky:2021VVCS} is in agreement with the results obtained with the applied deuteron potentials \cite{Hernandez:2014pwa}, we expect it to be consistent with \piEFT/.

As the total $2\gamma$ contribution we thus obtain: 
\begin{eqnarray}
    E_{2S}^{2\gamma} & = &E_{2S}^\mathrm{elastic}+E_{2S}^\mathrm{inel}+E_{2S}^\mathrm{hadr}+E_{2S}^\mathrm{eVP}+E_{2S}^\mathrm{Coulomb}\nonumber\\
    &=& -1.752(20) \text{ meV}.
\label{eq:total2g}
\end{eqnarray}
This agrees, within the uncertainty, with the recent compilation~\cite{Kalinowski:2018rmf}, $-1.740(21)$~meV. The earlier compilation~\cite{Krauth:2015nja} gave a smaller value, $-1.7096(200)$~meV, which, most importantly, missed the eVP correction, \Eqref{eVP}.

We now compare the theory prediction to the empirical extraction of the $2\gamma$-exchange contribution from the combined $\mu$D, $\mu$H, and H-D measurements. We start from the theory of the $\mu$D Lamb shift compiled in Ref.~\cite{Antognini:2012ofa}, together with an updated NLO hadronic vacuum polarization contribution \cite{Karshenboim:2021jsc}, see  Ref.~\cite{Antognini:2022xoo} for details\footnote{We have corrected a sign mistake in the inclusion of \cite[\#r8]{Antognini:2012ofa}. }:
\begin{eqnarray}
E_{2P-2S}(\mu{\rm D})   & = & 
\left[228.77408(38)-6.10801(28) \,\left(\frac{r_d}{\mathrm{fm}}\right)^2\right.\nonumber\\
& & - E_{2S}^{2\gamma}+0.00219(92)\bigg]\mathrm{meV}.
\label{eq:new_theory_LS}
\end{eqnarray}
The first term here contains QED and other structure-independent effects. The second one is the finite-size contribution. The last two are the nuclear-structure effects from, respectively, $2\gamma$ and 3$\gamma$-exchange.
For the latter, we are taking the sum of the elastic \cite[\#\#r3 and r3']{Antognini:2012ofa} and inelastic contributions \cite{Pachucki:2018yxe}. This updated $\mu$D theory is then compared with the experimental result from the CREMA collaboration~\cite{Pohl1:2016xoo},
\beq
E_{2P-2S}=202.8785(31)_\mathrm{stat}(14)_\mathrm{syst}\, \mathrm{meV},\eqlab{muDexp}
\eeq
and using the value of $r_d(\mu\text{H, iso})$ from Eq.~\eqref{eq:rdiso}, we obtain the following empirical extraction of the $2\gamma$-exchange contribution:
\beq
E _{2S}^{2\ga }(\text{emp.})=-1.7585(56)\,\mathrm{meV}.\eqlab{deuteronstrucNEWemp}
\eeq
The prediction obtained in the \piEFT/ framework, Eq.~\eqref{eq:total2g}, is fully consistent with this updated empirical result, but has an about $3.5$ times larger uncertainty.

Finally, using Eqs.~\eqref{eq:new_theory_LS},~\eqref{eq:muDexp}, and the prediction for $E_{2S}^{2\gamma}$, Eq.~\eqref{eq:total2g}, we obtain
\begin{equation}
r_d(\mu\text{D}) =2.12763 (13)_\text{exp}(77)_\text{theory}= 2.12763(78)~\text{fm}\label{eq:rdmuDOUR}
\end{equation}
for the deuteron charge radius extracted from $\mu$D Lamb shift, seen as ``N3LO pionless EFT'' in Fig.~\ref{fig:RdSummary}. Compared with the original extraction~\cite{Pohl1:2016xoo}, this value, although having a similar uncertainty, is in much better agreement with the combined H-D isotope shift and $\mu$H determinaiton, Eq.~\eqref{eq:rdiso}.

\section{Summary}\label{SecVI}
An accurate interpretation of the $2\gamma$-exchange correction to the $\mu$D Lamb shift is now possible, thanks to the unprecedented experimental precision achieved by the CREMA collaboration~\cite{Pohl1:2016xoo}. Here we employ a nuclear EFT which, among other good features, allows one to systematically improve the theoretical description of nuclear effects and to quantify the associated uncertainty.

We report on a calculation of the $2\gamma$-exchange contribution to the Lamb shift in $\mu$D, performed at N3LO in \piEFT/. Based on our earlier calculation of the deuteron VVCS amplitudes~\cite{Lensky:2021VVCS}, we calculate the elastic and inelastic  $2\gamma$-exchange contributions at $O(\alpha^5)$. The resulting elastic contribution shows a significant discrepancy with the data-driven result~\cite{Carlson:2013xea}, which uses the empirical deuteron form factors of Abbott et al.~\cite{Abbott:2000ak}. We argue that the latter parametrisation is not suitable for the low-$Q$ regime; the correlation between the deuteron charge and Friar radii can be used to judge the quality of the low-$Q$ description. The resulting inelastic contribution, on the other hand, is consistent with the recent evaluations~\cite{Acharya:2020bxf,Hernandez:2019zcm}. In addition, we evaluate the eVP corrections to the $2\gamma$-exchange contribution, for the first time in the EFT framework.

The systematic nature of the \piEFT/ expansion allows us to apply a Bayesian procedure to quantify the theoretical uncertainty along the lines of~Refs.~\cite{Furnstahl:2015rha,Perez:2015ufa} (in this context see also Ref.~\cite{LiMuli:2022kcy}, which applies Bayesian reasoning to study the uncertainties arising from a different expansion in muonic atoms and ions).

We also evaluate the higher-order contributions coming from the single-nucleon structure, which are expected to be the dominant terms beyond N3LO. This is achieved by departing from a strict \piEFT/ formalism; we believe that purely nuclear higher-order contributions play only a very minor r\^ole compared to these hadronic corrections.

We have updated the original empirical extraction of the $2\gamma$-exchange contribution~\cite{Pohl1:2016xoo}, resulting in a somewhat improved uncertainty. Our N3LO \piEFT/ prediction is fully consistent with it, but for now has a larger uncertainty. The same, in turn, is valid for the two alternative extractions of the deuteron charge radius, i.e., the value obtained from combining the $\mu$H Lamb shift with the H-D isotope shift \cite{Antognini:2012ofa}, and the one obtained from the $\mu$D Lamb shift using our \piEFT/ calculation, Eq.~\eqref{eq:rdmuDOUR}. Thus, the $\mu$H and $\mu$D extractions are, via the H-D isotope shift, consistent with each other; the original inconsistency was mainly caused by the missing eVP contribution, as already noted in~\cite{Kalinowski:2018rmf}.

As for an outlook, we note that, at the current level of precision, the single-nucleon effects are becoming appreciable, despite being a few orders of magnitude smaller than the nuclear effects. The nucleon structure is even more important in more tightly bound nuclei, such as helium. This emphasises the importance of investigating the nucleon structure, such as the nucleon polarisabilities. Furthermore, treating the nuclear and nucleon structure consistently, within the same EFT framework, would be an important step to an improved theoretical description.

\section*{Acknowledgements}
We thank A.~Hiller Blin  for useful communications and remarks on the manuscript. We thank V.~Baru and A.~Filin for sharing with us the details of their $\chi$ET calculation of the deuteron charge form factor. We thank C.~Carlson and M.~Gorchtein for useful communications. The calculations in this work were performed with the help of \textsc{FORM}~\cite{Vermaseren:2000nd}, and the figures in the article were made with the help of \textsc{SciDraw}~\cite{Caprio:2005dm}. We acknowledge the support of the Deutsche Forschungsgemeinschaft (DFG) through the Emmy Noether Programme [grant 449369623] and the Research Unit FOR5327 [grant 458854507]. F.H. acknowledges the support of the Swiss National Science Foundation (SNSF) through the Ambizione Grant PZ00P2\_193383.

\end{document}